\documentclass[conference]{IEEEtran}

\usepackage[utf8]{inputenc}
\usepackage{tikz}
\usepackage{amsmath, amssymb}
\usepackage{tabularx}
\usepackage{csquotes}
\usepackage[final,tracking = true,kerning = true,spacing = true]{microtype}
\usepackage[colorlinks = true, citecolor = blue]{hyperref}
\usepackage{subcaption}
\usepackage{booktabs}
\usepackage{acronym}
\usepackage{siunitx}

\microtypecontext{spacing = nonfrench}

\usepackage[style=ieee]{biblatex} 
\begin{document}

\date{}

\title{Learning automated defense strategies using graph-based cyber attack simulations}

\author{
\textrm{Jakob Nyberg}\\
KTH Royal Institute of Technology\\ 
Stockholm, Sweden\\
jaknyb@kth.se\\
\and
\textrm{Pontus Johnson}\\
KTH Royal Institute of Technology\\  
Stockholm, Sweden\\
pontusj@kth.se\\ 
}

\author{\IEEEauthorblockN{Jakob Nyberg}
\IEEEauthorblockA{\textit{KTH Royal Institute of Technology} \\
Stockholm, Sweden \\
jaknyb (at) kth.se}
\and
\IEEEauthorblockN{Pontus Johnson}
\IEEEauthorblockA{\textit{KTH Royal Institute of Technology} \\
Stockholm, Sweden \\ 
pontusj (at) kth.se}
}

\maketitle

\begin{abstract}
	We implemented and evaluated an automated cyber defense agent. The agent
	takes security alerts as input and uses \acl{RL} to learn a policy for
	executing predefined defensive measures. The defender policies were trained
	in an environment intended to simulate a cyber attack. In the simulation, an
	attacking agent attempts to capture targets in the environment, while the
	defender attempts to protect them by enabling defenses. The environment was
	modeled using attack graphs based on the \acl{MAL} language. We assumed that
	defensive measures have downtime costs, meaning that the defender agent
	was penalized for using them. We also assumed that the environment was
	equipped with an imperfect \acl{IDS} that occasionally produces erroneous
	alerts based on the environment state. To evaluate the setup, we trained the
	defensive agent with different volumes of \acl{IDS} noise. We also trained
	agents with different attacker strategies and graph sizes. In experiments,
	the defensive agent using policies trained with reinforcement learning
	outperformed agents using heuristic policies. Experiments also demonstrated
	that the policies could generalize across different attacker strategies.
	However, the performance of the learned policies decreased as the attack
	graphs increased in size.
\end{abstract}



\acrodef{IDS}{intrusion detection system}
\acrodef{RL}{reinforcement learning}
\acrodef{MAL}{Meta Attack Language}
\acrodef{ML}{machine learning}
\acrodef{MDP}{Markov decision process}
\acrodef{POMDP}{partially observable Markov decision process}
\acrodef{PPO}{Proximal Policy Optimization}
\acrodef{TTC}{time-to-compromise}
\acrodef{FPR}{false positive rate}
\acrodef{FNR}{false negative rate}
\acrodef{TPR}{true positive rate}
\acrodef{TNR}{true negative rate}
\acrodef{RNG}{random number generator}
\acrodef{SDN}{software defined network}

\newcommand\tripwire{tripwire}
\newcommand\AND{AND}
\newcommand\OR{OR}

\section{Introduction}\label{intro}

Cybersecurity is an important concern in our increasingly digital society.
Consequences of cyber crime can be severe, both on an individual and societal
level. Apart from managing personal finances and information, digital systems
are also used in maintaining infrastructure such as power grids, water supply,
and public transportation. The intricacies of these systems provide a large
attack surface for malicious actors to exploit. To cover this attack surface, we
believe that a combination of automation and human expertise is needed. Using
automated systems to handle rote tasks of the decision-making process can allow
human operators to focus on higher-level tasks that are yet difficult for a
computer to perform.

An autonomic system is a system that can manage itself, and adapt to changes in
its environment~\cite{Kephart2003}. A common reference model for autonomic
systems is the MAPE-K loop~\cite{Kephart2003}. The MAPE-K loop consists of four
steps: \emph{monitor}, \emph{analyze}, \emph{plan}, and \emph{execute}. An
autonomic agent monitors its environment through \emph{sensors}
provided by the system, and executes actions through
\emph{actuators}. 

In our implementation of the MAPE-K loop, the sensors are \ac{IDS} modules for
providing alert signals, and the actuators a set of predefined
defensive operations, such that can be defined in an \ac{SDN} controller or
host-based IDS system like Wazuh~\cite{Wazuh}.

To analyze the state and plan
actions, we use a policy function optimized using \ac{RL} that uses the \ac{IDS}
signals as input and selects a predefined defensive measure to execute. An issue
with \ac{RL}, especially when proceeded with the prefix \enquote{deep}, is the
need for large amounts of online data collection. Policies are learned by
continually observing and interacting with the environment. If running the
system is expensive or slow, this can be costly since training an agent
can require thousands of episodes of experience. One approach to solve this
issue is to train in a simulated environment. The simulator attempts to imitate
the inputs, outputs and dynamics of the real system. The learned policy function
can then be transferred to a real system with little or no additional training.
This is sometimes referred to as \emph{sim-to-real} transfer~\cite{Zhao2020}.
This work investigates the first half of the sim-to-real process, the simulation
and training. To simulate a cyberattack, we employ approaches from the field of
\emph{threat modeling}, which focuses on identifying and
analyzing threats to a system. Specifically, we use attack graphs produced using
the \ac{MAL}~\cite{johnson2018meta} to model the cyberattack process.

Although this work focuses on intrusion response, a major challenge in network
defense is the \emph{detection} of misuse or compromise. We eschew this issue by
assuming that an \ac{IDS} capable of detecting the ongoing attack is in place in
the system we wish to apply our agent in, the output of which is used as input
to the decision policy. In practice, the \ac{IDS} can be a signature-based
system, or on based on machine learning. We lighten our assumption by allowing
the \ac{IDS} to be imperfect, failing to register some events and falsely
reporting on others. A question of interest to us was how well the defender
agent could perform with, or in spite of, the imperfect information from the
\ac{IDS}. We investigated this by training the agent with different volumes of
\ac{IDS} errors, and comparing the performance of the agent against heuristic
baselines. We were also interested in the ability of the \ac{RL} policies to
generalize to different attacker strategies. This was investigated by training
policies using \ac{RL} on a number of different attacker strategies, and
evaluating the performance of the agent on previously unseen ones. Lastly, we
investigated the effect of the size of the attack graph on the performance of
the \ac{RL} policies.

\section{Related Work}\label{related}

Cyber intrusion detection and response are broad research fields whose scope
extends far beyond that of this work. For the purpose of brevity, we focus this
section on other works that employ reinforcement learning for intrusion
\emph{response}. 

There have been several works that implement different varieties of automated
defender agents based on \ac{RL}~\cite{9596578, GabirondoLopez2021, Han2018, Hammar2021,
Seifert2020, Hu2020, Huang2022a}.
\Textcite{Huang2022a} list several applications where \ac{RL} can or may be used to
develop a cyber-resilient system. \Textcite{GabirondoLopez2021} trains a variant of
AlphaZero using self-play with an attacker and defender agent. \Textcite{Hammar2021}
presents and tries to solve the defense problem as an optimal stopping problem.
\Textcite{Hu2020} formulates the cyber environment as a \ac{POMDP} and uses Q-learning
to find a solution policy. Several of these works also implement their own cyber
attack simulators in order to learn defender policies. Named simulator include
CybOrg~\cite{MaxwellStanden2021}, CyberBattleSim~\cite{Seifert2020} and Yawning
Titan~\cite{Andrew2022}.

The work of~\textcite{Andrew2022} bears close resemblance this one. They, too,
present a graph-based cyberattack simulator and train a defender agent using the
output of the simulator. Unlike this work, however, the authors use causal
Bayesian optimization to train a policy function, rather than reinforcement
learning. This has the advantage of producing a causal model of the defender's
behavior. The causal model can then be used to interpret the defender's
decisions, which can be difficult with methods based on neural networks.
However, as noted by the authors, using causal optimization methods requires the
manual construction of a causal system model. This is a potentially difficult
and time-consuming process that may not be feasible for large or complex systems
with many variables that influence decision-making. The same attack simulator is
also used by~\textcite{Collyer2022}, but with an agent based on \ac{RL} instead of a
causal models. The authors use a graph embedding method to represent the graph,
and use a graph convolutional network to process the graph. This allows the
agent to generalize to graphs other than the one it was trained on. They note
that for large graphs, the embedding method leads to improved performance in
environments not seen by the agent during training.

The work of \textcite{Wolk2022} also share several aspects of this project. They
employ a variety of methods based on \ac{RL} and \ac{PPO} to train an automated defender
agent on the CAGE Challenge~\cite{cage_challenge_2}, which is built on top of the CybORG
simulator~\cite{MaxwellStanden2021} cyber environment simulator. They find that 
an approach using an ensemble of \ac{PPO} agents performed better than other
variants, such as hierarchical \ac{PPO}. They also evaluate their agents
in unseen environments, such as to test how well they may perform in a Real
environment, and find that the performance of drops significantly.

Related to the topic of automated defense agents is the concept of automated
penetration testing, that instead aims to automate the process of finding and
exploiting vulnerabilities in a system~\cite{Holm2022,
Gangupantulu2022, Ghanem2019, 9951035, Hu2020a}. Automated attacker agent can also serve
as opponents for automated defense agents, using adversarial machine learning or
game-theoretic solution methods. For this work, we use heuristic policies for
the attacker agent, but we believe that the use of \ac{RL} for the attacker
agent is a promising direction for future work.
\section{Reinforcement Learning}\label{sec:background}

Reinforcement learning is a field of machine learning focused on the interaction
between an \emph{agent} and an \emph{environment}~\cite[Ch. 1]{Sutton2018}. The
environment is usually modeled as a \ac{MDP}, and the agent should learn a
policy for performing actions that maximizes a reward signal. Unlike \ac{MDP}
solver methods such as dynamic programming, reinforcement learning does not
necessarily require a model of the environment. Instead, the agent interacts
with the environment and learns a policy from observations and rewards produced
by said environment. There are a number of different algorithms for
reinforcement learning, including Q-learning~\cite[Ch. 6]{Sutton2018} and policy
gradient methods~\cite[Ch. 13]{Sutton2018}. Policy gradient methods are focused
on directly finding a policy function through gradient descent. The policy
function maps a given state \(s\) to an action \(a\) via the parameters
\(\theta\), and is denoted as \(\pi_\theta(a|s)\). \ac{PPO} is a policy
gradient method that applies a clip function on the policy loss to limit the
amount of change in the policy parameters in a single
iteration~\cite{Schulman2017}. This is intended to improve the stability of the
learning process. The clipped policy loss is defined as

\begin{align}
	& L_t^{CLIP}(\theta) =\\
	& \hat{E_t} \left[ \min \left(r_t(\theta)\hat{A}_t, \text{clip}(r_t(\theta), 1-\epsilon, 1+\epsilon) \hat{A}_t \right) \right]
\end{align}

where \(r_t(\theta) = \frac{\pi_\theta(a|s)}{\pi_{\theta_t}(a|s)}\) is the ratio of 
the new policy to the old policy, \(\epsilon\) a hyperparameter that controls
the clip limits and \(\hat{A}_t\) is an advantage function.

\section{System Description}\label{sec:environment} The basis of the system
simulation is an \emph{attack graph}. An attack graph is a model to predict
and analyze possible events and outcomes of a cyberattack. The attack graphs
used are based on those produced using the \acf{MAL}~\cite{johnson2018meta},
which can be used to create digital twins of cyber-physical systems and
networks~\cite{Masi2023}. The graphs defined here share the logic and features
of \ac{MAL} attack graphs. Their construction, however, differs due to not being
generated using a list of defined object classes and relations, but rather are
constructed manually.

We define an attack graph as a directed graph \(G = (V, E)\), where \(V\) is the
set of nodes, and \(E\) is the set of edges between nodes. The nodes are divided
into two sets: \emph{attack steps}, \(A\), and \emph{defense steps}, \(D\). Each
attack step features a probability distribution specifying the time they require
to be \emph{compromised}, the \ac{TTC}. Defense steps, on the other hand, are
\emph{enabled}. When a defense step is enabled, it will prevent any attack steps
that has the defense step as a parent node from being compromised. A defense
step can represent a file not being encrypted, or a firewall rule not existing.
Activating the defense, e.g.\ making the file encrypted, makes the file no
longer readable.

When an attack step is compromised, it may give access to other steps in the
graph. For example, one attack step may represent the action of performing a
password dictionary attack. If the attack is successful, the attack step is
compromised, and the attacker may proceed to a next step in the graph.

\begin{figure}[tbp]
	\centering
	
\begin{subfigure}
		{0.45\textwidth}
		\centering
		\includegraphics[width = 0.7\textwidth]{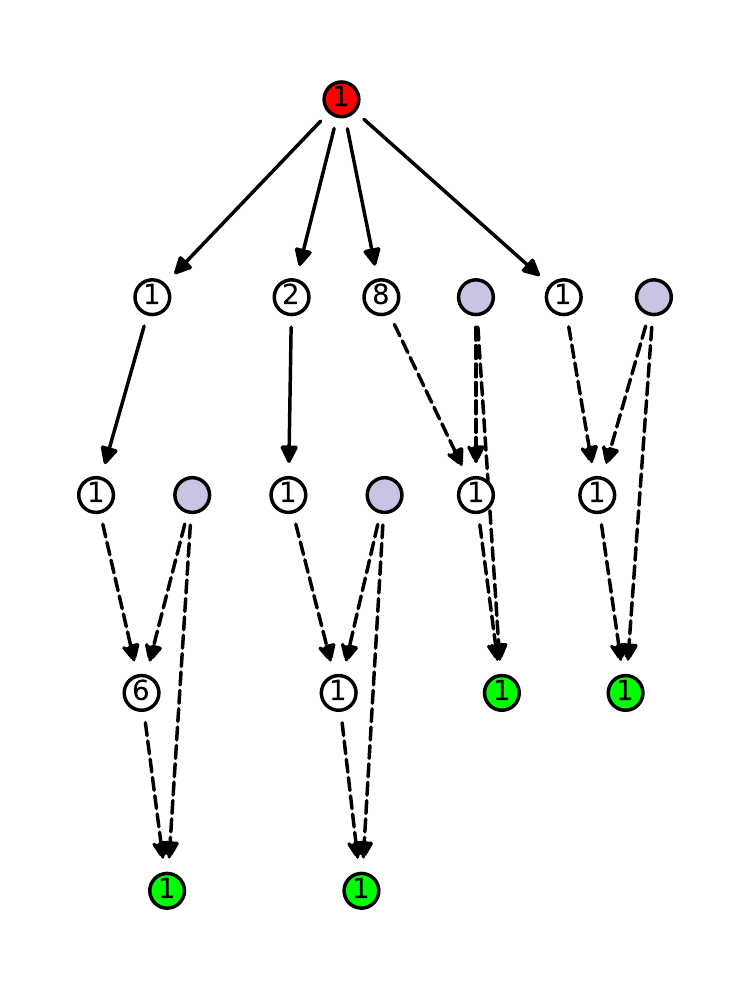}
		\caption{Attack graph used for \ac{FPR}/\ac{FNR} sweep experiment. Each of the four
			flags can be defended using a defense step.}\label{fig:4ways}
\end{subfigure}
	
\begin{subfigure}
		{0.45\textwidth}
		\centering
		\includegraphics[width = 0.7\textwidth]{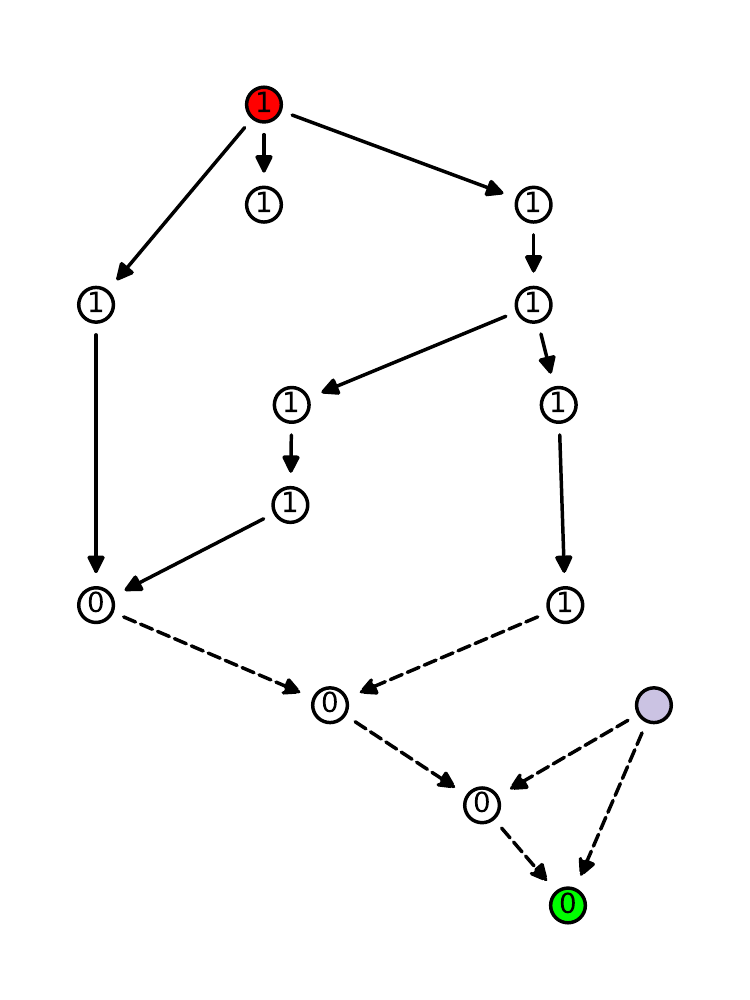}
		\caption{Attack graph used for attacker policy comparison experiment.
}\label{fig:2keys1door} 
\end{subfigure}
\caption{Attack graphs constructed manually for
		experiments. Defense steps are colored purple, flags green and attacker entry
		points red. AND-steps have dashed incoming edged. Average \ac{TTC}
		values are drawn as numbers in nodes.}\label{fig:graphs}
\end{figure}

In accordance to \ac{MAL}, attack steps can be of one of two types: \AND{}
and \OR. \AND-steps require all parent steps to be compromised in order to be
compromised themselves, whereas \OR-steps require only one parent step to be
compromised. \AND{}-steps can be used to represent preconditions such as an account
requiring a password to be compromised. \OR{}-steps, on the other hand, can be
used to represent alternative paths, such as a user account being compromised by
either a password dictionary attack or a phishing campaign. Figures~\ref{fig:2keys1door}
and~\ref{fig:4ways} show visualizations of handcrafted attack graphs used for
experimentation. \AND{}-steps are indicated by dashed incoming edges.

We define the \emph{attack surface} as a set of attack steps, \(A_{as} \subset A\),
that fulfill the following conditions:
\begin{enumerate}
	\item There is an edge from a compromised attack step to the step.
	\item One parent step is compromised, if the step is an \OR{}-step, or all parent steps are compromised, if the step is an \AND{}-step.
	\item A defense step that is a parent node to the attack step is not enabled.
\end{enumerate}

\subsection{Episode Structure}\label{sec:episode} In order to train a defender
agent, we model an attack-defend capture-the-flag games as an \ac{MDP}. The goal
of the attacker is to capture as many targets, \emph{flags}, as possible. The
defender attempts to stop the attacker, and is penalized for each flag
compromised.

The game is played with an attack graph acting as the basis of actions that can
be taken by the attacker and defender. A subset of attack steps in the graph
are assigned as flags, \(F \in A\). Episodes are initiated with a single attack step in the
attack graph being compromised, the entry point for the attacker.

The game is played in discrete time-steps, with the attacker and defender both acting within
the same time-step. Every time-step, the attacker can select any attack step from the
attack surface to work on. For every time-step the attacker works on an attack step, the
\ac{TTC} value is reduced by 1. When the \ac{TTC} reaches 0, the attack step is
compromised. As the defender's set of actions, it can select any defense step
from the set of defense steps that are not enabled. When a defense
step is enabled, all child steps are removed from the attack surface and can
no longer be compromised by the attacker. Steps that have been compromised by
the attacker are no longer considered compromised if defended. The defender can
also choose to do nothing for a time-step. An episode ends when the attack surface is
empty, meaning that there are no attack steps that can be reached by the
attacker.

\subsection{State Description}\label{sec:state}

The state of the environment is expressed by two vectors, \(\vec{A} =
(a_0,\ldots, a_{|A|}) \in \left\{0, 1\right\}^{|A|}\) and \(\vec{D} = (d_0,\ldots, d_{|D|}) \in \left\{0,
1\right\}^{|D|}\). \(\vec{A}\) describes the state of all attack steps, and
\(\vec{D}\) the state of all defense steps. For \(\vec{A}\), 1 represents that
the step is compromised, and 0 that it is not. For \(\vec{D}\), 1 represents
that the defense is enabled, and 0 that it is disabled.

We assume that the network that is being defended has an \ac{IDS} capable of
tracking the state of every attack step in the attack graph. We also assume that
this process may be faulty, and that the \ac{IDS} may sometimes fail to report an
attack step as compromised, or report a step as compromised when it is actually not.
Thus, we define another vector, \(\vec{O} = (o_0\ldots,o_{|A|}) \in \{0, 1\}^{|A|}\), to
describe the observation of the attack steps produced by the \ac{IDS}. As the
state changes over time, the subscript \(t\) is used to denote the state at time-step
\(t\), e.g.\ \(\vec{A}_t\) is the state of all attack steps at time-step \(t\).

We define the accuracy of the \ac{IDS} by a \ac{FPR} and \ac{FNR}. Every time-step,
depending on the \ac{FPR} and \ac{FNR}, the state of the system can be
incorrectly reported by the \ac{IDS}. The \ac{FPR} and \ac{FNR} can be expressed
as the conditional probabilities
\begin{align}
	 & \text{FPR} = P(o_{ti} = 1|a_{ti} = 0) & i \in \{0, \ldots, |A|\} \\
	 & \text{FNR} = P(o_{ti} = 0|a_{ti} = 1) & i \in \{0, \ldots, |A|\}
\end{align}

\subsection{Attacker}\label{sec:attackers}

The attacker agent uses a policy function to select actions defined by the
attack graph, deciding which attack step to work on at a given time-step. 
The attacker agent can only select attack steps that are in the attack surface.
Four different search policies were used in experiments for the attacker agent.

\paragraph{Random} A search policy that selects a random attack step from the
attack surface to work on each time-step.

\paragraph{Breadth-first} 
A search policy that compromises all steps on the current depth of the attack
surface before moving on to the next depth. The policy is randomized by shuffling
the order that attack steps are traversed when multiple are 
available at the same depth.

\paragraph{Depth-first}
A search policy that compromises each branch of the attack surface until it
reaches an attack step with no children, and then backtracks to the start of another
branch. The policy is randomized by shuffling the order in which branches are
traversed when multiple are available.

\paragraph{Pathfinder} A policy that incorporates full information of the attack
graph. It calculates the shortest path to each flag, and then targets flags in
order of increasing \ac{TTC} cost for the path. If blocked by a defense step en route,
the policy will recalculate the shortest path to the targeted flag. If
no path is available, it will target the next flag in the list.

\subsection{Defender}

The defender agent takes alert signals from the \ac{IDS} as input, and makes a
decision on which defense to enable. It can also choose to do nothing for a
time-step. Decisions are made using a policy function, with \(\vec{A}\)
concatenated with \(\vec{D}\) as input. The action space consists of all defense
steps in \(D\) that are not enabled, plus a do-nothing action. In experiments, 
we evaluated three choices of defender policy function:

\paragraph{Random} A policy that selects an available defense step at random
each time-step.

\paragraph{Tripwire} A conditional policy, that emulates the behavior of
\enquote{if-this-then-that} rules present in \ac{IDS} frameworks such as
Wazuh~\cite{Wazuh}. An example rule may be to automatically block traffic from a
certain host if a port scan is detected. Within the attack graph environment,
this is translated to activating a defense step when one of its child step is
reported as compromised. 

\paragraph{Reinforcement Learning}
A policy trained using \ac{RL}. The policy function was parametrized by a
fully-connected neural network and optimized using the \ac{PPO} algorithm. The
neural network had an input layer of size \(|A| + |D|\), a set number of hidden
layers, and an output layer of size \(|D| + 1\). The number of hidden layers and
their sizes were treated as hyperparameters. \(\tanh\) was used as the
activation function for the hidden layers.

\subsection{Reward}

The defender agent has two goals in the game: to minimize the number of compromised
flags, and to minimize the operational cost of the defense. The operational cost
is an abstraction that could, for example, represent the cost of downtime
resulting from disabling a service or network for security purposes. A defender
with the singular goal of defending the flags could easily maximize its reward
by shutting off access to the entire system immediately and be done, as the
attacker will have nothing to do at that point. However, this would prevent
normal users from accessing the system, which we consider undesirable. Defensive
measures are thus assumed to have a set cost \(c_d\). With \(\vec{F_t}=\{f_0,
\ldots f_{|F|}\} \in \{0, 1\}^{|F|}\) denoting flag states, the reward for the
defender agent at a given time-step \(t\) is defined as
\begin{align}
	 & r_d(t) = -\sum_{i=0}^{|D|} d_ic_{d} - \sum_{i=0}^{|F|}f_ic_{f} & d_i, f_i \in [0, 1],\ c \in \mathbb{R}
\end{align}
where \(c_d\) is the cost of activating a defense step \(d \in D\), \(c_f\) is
the cost of taking a flag \(f \in F\). Note that \(f_i=1\) only if the flag was
taken at time-step \(t\), meaning that the penalty is only incurred once for
each flag taken. The defense penalty, on the other hand, is incurred on every
time-step for every defense step enabled, i.e. \(d_i=1\). There are no
positive terms in the reward function, making the maximum possible reward 0. The
minimum reward would be incurred if the defender enables all defenses from the
first time-step, and the attacker still takes all flags during the episode. As
the defender agent can only disable one defense per time-step, the minimum
cumulative reward for an episode of length \(l\) is
\begin{align}
	-c_d\left(\sum_{i=1}^{|D|-1} i + |D|(l-\left(|D|-1\right))\right) - c_f|F|
\end{align}

\section{Experiments}\label{sec:experiments}

Three experiments were performed to evaluate the simulator and defender agent: A
comparison between the \ac{RL} policy and heuristic policies at different levels
of \ac{IDS} accuracy, a comparison between \ac{RL} policies trained with
different attackers and an analysis of how the policy training is affected as
the graph size increases.

\ac{RL} policy training and evaluation was performed on a Google Cloud virtual
machine equipped with an Nvidia V100 GPU, 12 CPU cores, and 30 GB of RAM\@. The
implementation used the Python library Ray \ac{RL}Lib~\cite{Liang2018} for 
implementations of the reinforcement learning algorithms. All
policies trained with \ac{RL} were run for 500 \ac{PPO} policy iterations. All
policies were evaluated by running 500 episodes. Each policy took roughly
20 minutes to train and evaluate.

Experiments were performed
three times with different seeds for the random number generator. The neural
networks used two hidden layers with 128 nodes.

All \ac{TTC} values except those explicitly set to 0 were initialized at 
the start of each episode by sampling from the exponential distribution \(f(x; \beta) =
\frac{1}{\beta} e^{-{\frac{1}{\beta}} x}\), where \(\beta\) is the mean \ac{TTC}
value assigned to each attack step. 

The costs were set as \(c_f = 1.5 \cdot \sum_{a\in A} \ac{TTC}(a) \) and \(c_d =
1\) for all flags and defense steps respectively, where \(\ac{TTC}(a)\) denotes
the \ac{TTC} value for attack step \(a\). The choice of \(c_f\) was made such
that the cost of losing a flag would be greater than disabling a defense step on
the first time-step.

\newcommand\firstexperiment{Sensor Fault Resistance}
\subsection{\firstexperiment}

A desired property of a good defender agent is the ability to operate in spite
of imperfect information. Therefore, an experiment was performed to study the
effects of false alerts and missed alerts on the defender. The rate of errors in
the observations is defined using two values, the \ac{FPR} and the \ac{FNR}.
Five values for the \ac{FPR} and \ac{FNR} were selected: \SI{0}{\%},
\SI{12.5}{\%}, \SI{25}{\%}, \SI{72.5}{\%} and \SI{100}{\%}. A partial grid
search was performed over combinations the error rates, for a total of 15
points. Points where \(FNR > 1 - TNR\) were excluded since they are equivalent
to those where \(FNR < 1 - TNR\), but with reverse definitions of the true and
false labels.

Three defender agents, one using policies trained with \ac{PPO}, one using the
\tripwire{} policy and one using the random policy were compared.
\autoref{fig:4ways} depicts the attack graph used as the environment, with
depth-first search as the attacker policy. For the \ac{RL} agents, one policy
was trained and evaluated for each combination of \ac{FPR} and \ac{FNR} values.

\newcommand\secondexperiment{Attacker Comparison}
\subsection{\secondexperiment}\label{sec:attackercomparison}

A second experiment was performed to study the generalization capabilities of
the defender agent trained using \ac{RL}. As such, only the \ac{PPO} policy was
used for this experiment. The agent was trained with one attacker policy, and
then evaluated against all attacker policies listed in
Section~\ref{sec:attackers}. A mixture of the attacker policies was used as an
additional variant, where the attacker policy was chosen at random at the
beginning of each episode. \autoref{fig:2keys1door} depicts the attack graph
used. The \ac{FPR} and \ac{FNR} were set to \SI{10}{\%}.

\newcommand\thirdexperiment{Graph Size}
\subsection{\thirdexperiment}
Real-life attack graphs can be huge due to the complexity of real-life
systems~\cite{johnson2018meta}. Thus, an experiment was performed to study the
effects of graph size on policy learning.

Four graphs of sizes 20, 40, 60 and 80 steps were generated using a semi-random
attachment procedure. Each graph has \(\frac{|A|}{20}\) flags. Each flag step
had a defense step attached to it, ensuring that each flag was defensible. A
policy was trained for each graph size using \ac{PPO} and evaluated together
with the \tripwire{} policy. The same hyperparameter values were used for
training on all graph sizes. The \ac{FPR} and \ac{FNR} were set to \SI{10}{\%}.

\newpage
\section{Results}\label{sec:results}

Sections~\ref{sec:sensornoise},~\ref{sec:attackerresults}
and~\ref{sec:trainingtime} present the results of the experiments described in
Section~\ref{sec:experiments}. Results for all experiments are averaged over
three runs with different seeds for the \ac{RNG}. The \ac{RNG} affects the
\ac{TTC} values sampled at that episode initialization, attacker behavior, and
neural network parameter initialization.

\subsection{\firstexperiment}\label{sec:sensornoise}

\begin{figure}[!htpb]
	\centering

\begin{subfigure}
		{0.48\textwidth}
		\centering
		\includegraphics[width = \textwidth, trim = {0, 3cm, 0, 4cm}, clip]{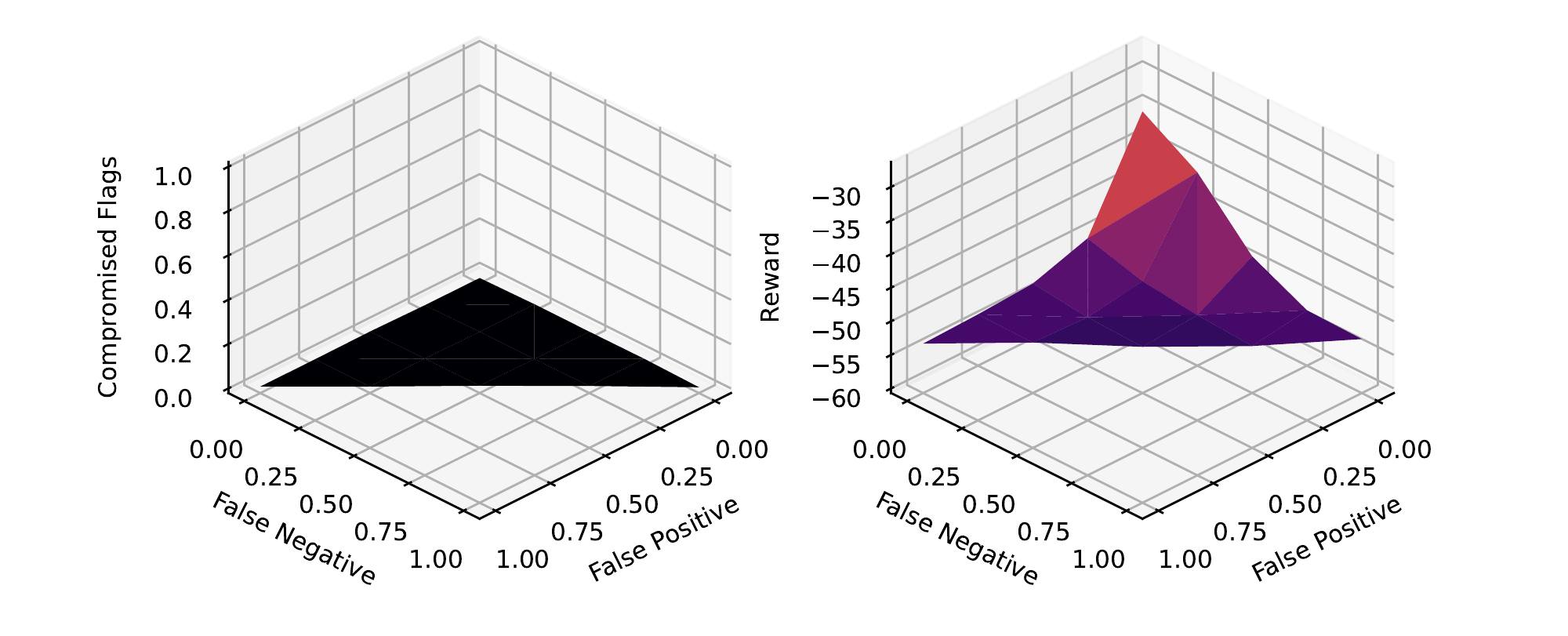}
		\caption{Surface plots showing performance metrics for defender agent
		using \ac{RL} policies. The percentage of flags taken by the attacker remains low
			for all values of the \ac{FPR} and \ac{FNR}.}\label{fig:surface_ppo}
\end{subfigure}
	
\begin{subfigure}
		{0.48\textwidth}
		\centering
		\includegraphics[width = \textwidth, trim = {0, 3cm, 0, 4cm}, clip]{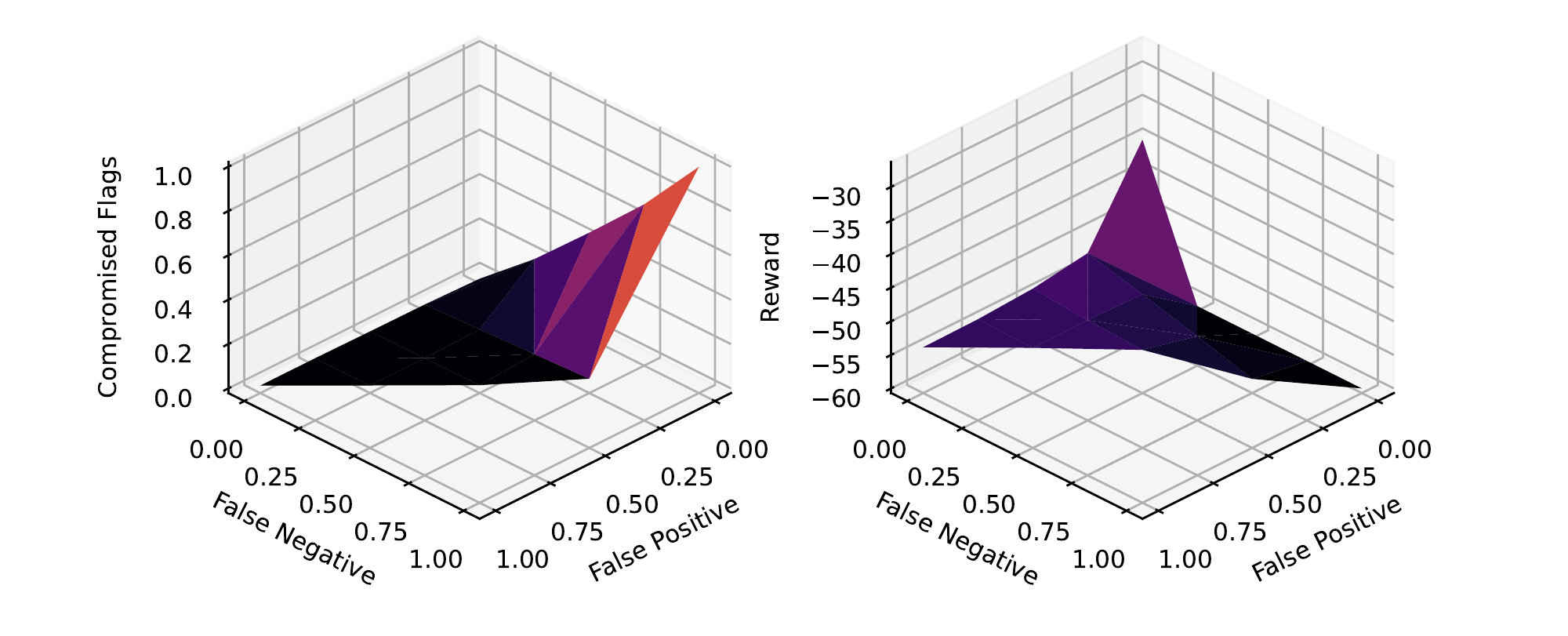}
		\caption{Surface plots showing performance metrics for defender using the
			\tripwire{} policy. As the \ac{FNR} increases, performance decreases
			significantly.}\label{fig:surface_tripwire}
\end{subfigure}
	
\begin{subfigure}
		{0.48\textwidth}
		\centering
		\includegraphics[width = \textwidth, trim = {0, 3cm, 0, 4cm}, clip]{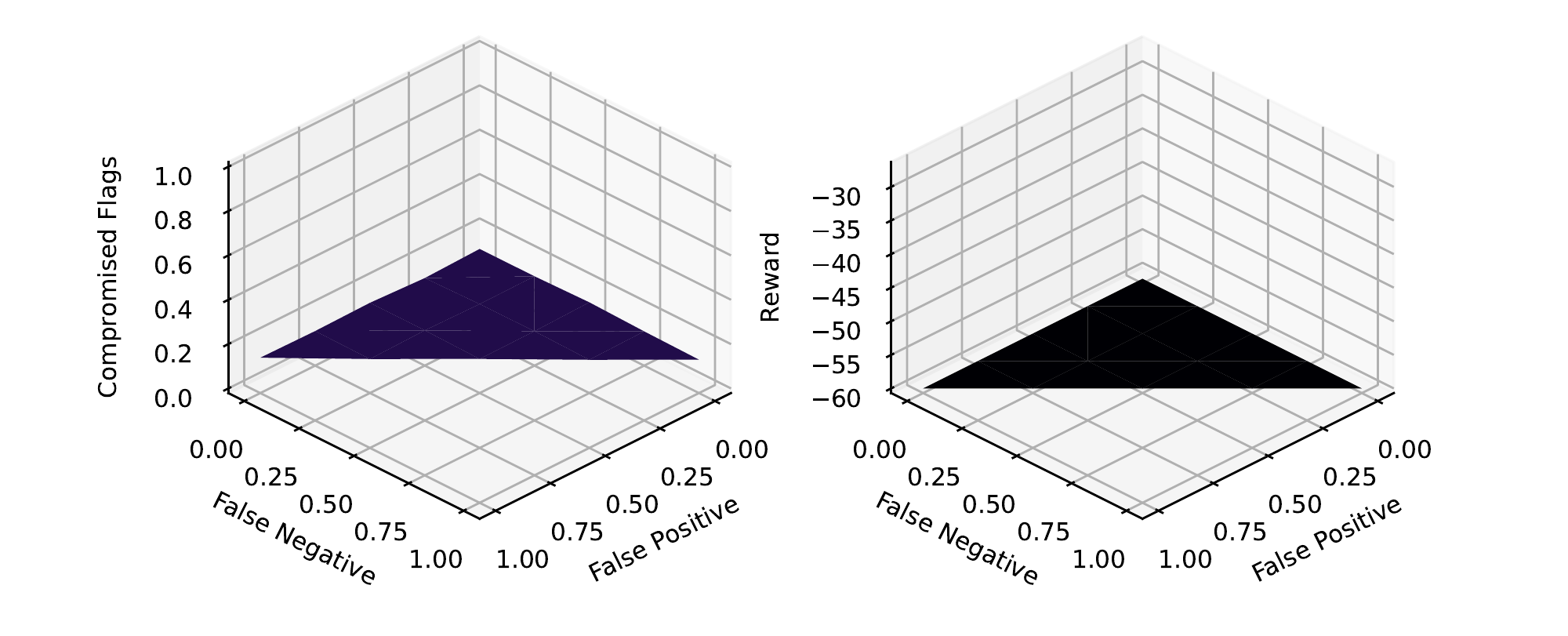}
		\caption{Surface plots showing performance metrics for defender taking
			random actions. The reward is consistently low compared to other
			methods, and the amount of flags lost to the attacker is
			higher.}\label{fig:surface_random}
\end{subfigure}

\caption{Surface plots showing metrics for different defender policies.
		Results are averages over three runs with different \ac{RNG}
		seeds.}\label{fig:surface_plots}
\end{figure}

\autoref{fig:surface_plots} presents performance metrics for the three defender
agents as surface plots. There are two figures for each policy, one showing the
average percentage of flags captured by the attacking agent, and one showing the
average reward for the 500 episodes of evaluation. Episodes were, in average, 13
time-steps long for agents using \ac{PPO} and \tripwire{} policies, and 15
time-steps for the agent using a random policy. The agent using the random
policy has a consistent performance across the different combinations, as it
does not use observations to begin with. The amount of flags captured was higher
than for the other policies, and the reward was thus lower.

For all but two combinations of error rates, the learned policies produce a
higher average reward than the \tripwire{} policy. For the two combinations
where the \(\ac{FPR}\geq0.5\) and the \(\ac{FNR} = 0\), the difference in reward
was no longer statistically significant (\(p<0.05\)).

The performance of both learned polices and the \tripwire{} policy decrease as
the \ac{FPR} increases. The \ac{PPO} policies performs better when the \ac{FNR}
increases compared to the \tripwire{} policy, the performance of which rapidly
decreases with an increased \ac{FNR}. 

Averaged across all combinations, the \ac{PPO} policies have an average
cumulative reward of \(-50\), compared with \(-73\) for the \tripwire{} policy
and \(-72\) for the random policy.

\subsection{\secondexperiment}\label{sec:attackerresults}

\begin{figure}
	\centering
	\includegraphics[width = 0.5\textwidth]{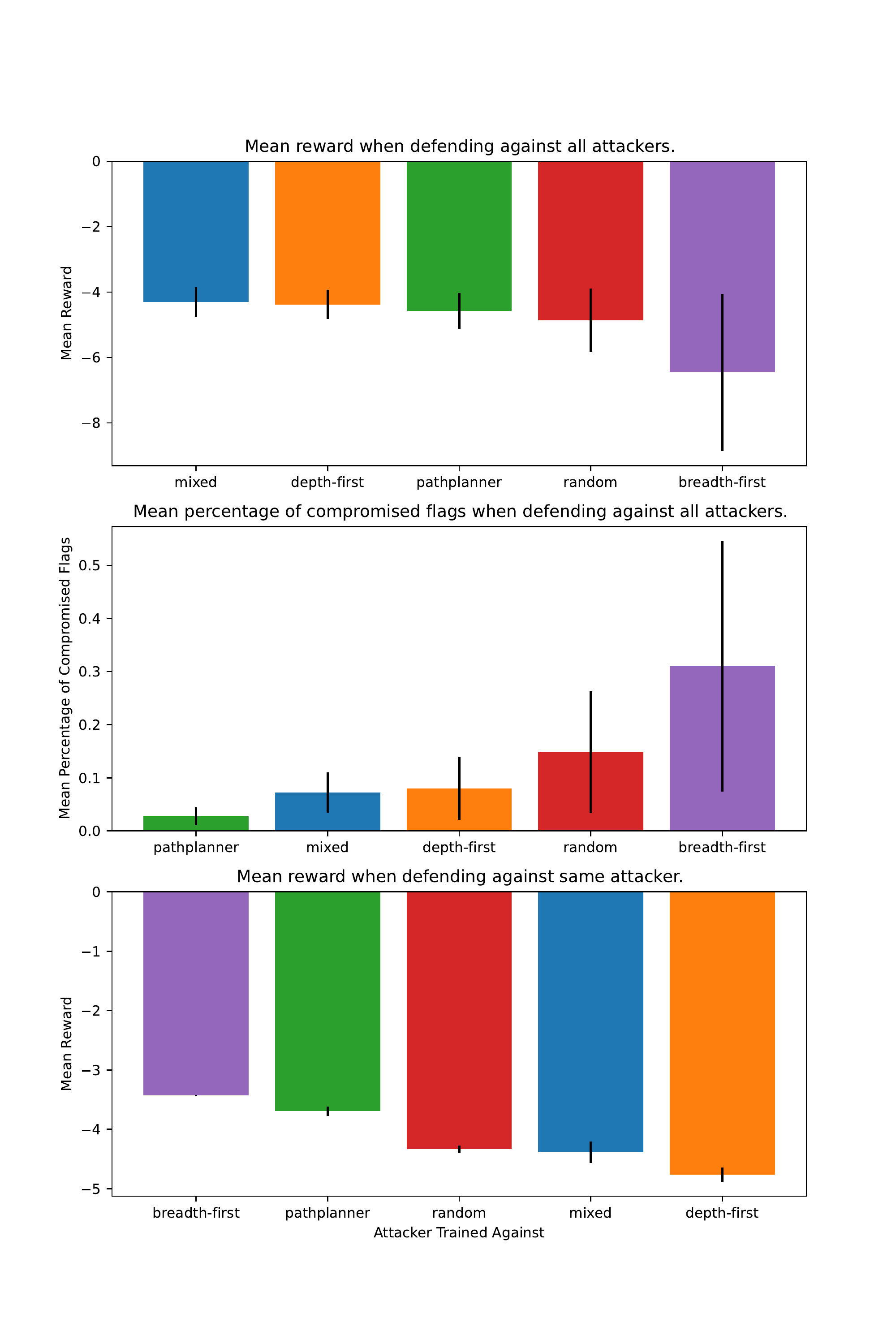}
	\caption{Three plots showing \ac{PPO} policy performance against different
	attacker policies during evaluation. The top and middle plots show the
	average reward and flags captured when faced with attacker policies not
	encountered during training. The bottom plot shows the reward when evaluated
	against the attacker policy used during training. The metrics are averaged
	over three runs with different \ac{RNG} seeds.}\label{fig:ppo_bar}
\end{figure}

\autoref{fig:ppo_bar} shows bar plots of the rewards and percentage of flags
compromised. Episodes were, in average, 9 time-steps long. The longest
recorded episode was 19 time-steps, and the shortest 6 time-steps.

There is a significant difference between the rewards produced by the agents
when faced with different attacker policies than those they were trained on. The
policy trained against the depth-first attacker produces the lowest reward
compared to the others when faced with the depth-first attacker. However, it
performs much better on average when faced with other strategies. Inversely, the
policy trained against the breadth-first attacker has the best overall
performance compared to the other policies when faced with only one attacker,
but is significantly worse when faced with others. When averaged over all 
attackers, the difference in reward between policies not trained against the
breadth-first attacker were not statistically significant (\(p<0.05\)).

\subsection{\thirdexperiment}\label{sec:trainingtime}

The performance of the policies learned by \ac{PPO} decrease as the graph size increases.
For all graph sizes greater than 20, the \ac{RL} policies produced worse results
than the \tripwire{} policy. The \tripwire{} policy also decreases in
performance as the graph size increases, but the decrease is less than for the
\ac{RL} policies.

\section{Discussion}\label{sec:discussion}

\subsection{Other Baselines}

We compared the \ac{RL} policy against two heuristic policies, neither of which
were optimal. Comparing against the optimal policy would be beneficial to judge
the effectiveness of the learned policies. In order to find the optimal policy
for a particular configuration, one could apply \ac{POMDP} solver methods. These
may become difficult to apply, however, for larger graphs as the number of
possible states is \(2^{n}\), two to the power of attack steps. The state space
thus grows exponentially. This also makes it difficult to compare the neural
network policy to a tabular \ac{RL} policy, like Q-learning that store a value
for each state-action pair.

\subsection{Scaling}

A large drawback of the \ac{RL} method was the time it took to train the neural
network policy function. The training time only grew larger as graphs sizes
increased and, inversely, the performance of the policies decreased. The
decrease in performance with increased graph sizes comes from a number of
factors. One of these was optimization robustness. As the graph size increases,
the size of the neural network grows linearly. Despite the size of neural
network and reward signals changing, the same set of hyperparameters was used
for all graph sizes. A procedure to find appropriate hyperparameters for each
graph size would likely be necessary to make training more robust. One could
also take an orthogonal approach and apply same neural network to all graph
sizes. This would require a different approach to how the graph state was
represented, as the current approach is not invariant to changes in the graph
structure or size. Graph convolutional networks are a possible solution to this
problem, as demonstrated in previous work~\cite{Collyer2022}.

Another complicating factor was the rules of the attack-defense game. When the
graph size increases, so does the episode length. The episode ended only when
the attacker no longer can perform any actions. Even if the defender enables all
defenses, the attacker can still traverse the remaining nodes, thus extending
the episode length. Adjusting the episode end conditions could help alleviate
this issue, such as imposing a time limit for the attacker. Another issue that
comes from long episodes is that the \ac{PPO} policies are stochastic. This
means that the longer the episode, and the more actions are available, the more
likely it is that the defender will eventually take a low-probability action.
As there is no way to undo actions, the defender can eventually enable every
defense even if the attacker does nothing. This could be addressed by adding a
probability threshold for enabling a defense, or by always selecting the most
likely one.

\subsection{Generalization}

Generalization is a desirable property for the defender agent, as it would allow
the agent to be used in different environments without having to retrain the
policy. This was also one of the main motivations for using deep \ac{RL}, as we
would like to be able to learn policies in environments that are easily
simulated, in order to be able to test them in more realistic environments down
the line. \autoref{fig:ppo_bar} demonstrates that the policy can generalize
across different attacker policies. However, so can the \tripwire{} policy, as
it is invariant to the behavior of the attacker. A more interesting task is to
generalize to different graph environments. In the current implementation, a
learned policy function is only applicable to a single graph. The graph topology
has to be static, and no attack steps can be added or removed. Additionally, the
agent was only presented with the node states, and not the topology of the
graph. Both of these problems may be addressed by using a model that can
incorporate the graph topology, like a graph neural network. Another task to
test generalization is to evaluate trained policies against different volumes of
\ac{IDS} noise, to determine which level is best to train the policy on.

\subsection{The Sim-to-Real Gap}

The defender agent is trained in a simulated environment, with the intention of
being used in a real environment. Within the field of robotics, it has been
shown that the transfer of learned policy functions from simulation to real
world can deteriorate performance of the policy~\cite{Zhao2020}. This is a
consequence of differences between the real world and the simulation model,
leading to worse policy performance. For this work, the target environment is a
computer network, and we should expect that the performance of the learned
policy function will change when applied to real alert patterns. Measuring the
magnitude of this change will be a key focus for future work. One issue related
to the field of cybersecurity in particular is the construction of realistic
attack scenarios. Even if the network itself is modeled accurately, the attacks
and adversaries the defender trains against may be unrealistic. 

\section{Conclusion}\label{sec:conclusions}

An automated cyber defense agent using policies trained with \ac{RL} was
implemented and evaluated. The agent was trained and tested in a simulated
environment based on \ac{MAL} attack graphs. The \ac{RL} policy was compared
against heuristic policies when faced with different volumes of \ac{IDS} noise.
We observed that the agents implemented using \ac{RL} could learn policies that
were better than the heuristic policies, and that they were more resilient to
missed alerts. From our second experiment, we also observed that the neural
network-based policy had the ability to generalize to different adversarial
attackers, and that some adversaries were better training opponents than others.
Unfortunately, the performance of the learned policies decreased as the graphs
increased in size. Future work will focus on the ability for the defender agent
to generalize across graphs, and crossing the gap between simulation and the
real world, transferring a defender agent trained in simulation to an actual
computer network.

\printbibliography%

\newpage

\end{document}